\documentclass{webofc}
\usepackage[varg]{txfonts}   
\usepackage{natbib}
\begin{document}
\title{Using CMS Open Data in research -- challenges and directions}
%
%

\author{\firstname{Kati} \lastname{Lassila-Perini}\inst{1}\fnsep\thanks{\email{k.lassila-perini@cern.ch}} \and
        \firstname{Clemens} \lastname{Lange}\inst{2}\fnsep\thanks{\email{clemens.lange@cern.ch}} \and
        \firstname{Edgar} \lastname{Carrera Jarrin}\inst{3}\fnsep\thanks{\email{ecarrera@usfq.edu.ec}} \and
        \firstname{Matthew} \lastname{Bellis}\inst{4}\fnsep\thanks{\email{mbellis@siena.edu}}
}

\institute{Helsinki Institute of Physics, Finland 
\and
           CERN, Switzerland 
\and
           Universidad San Francisco de Quito, Ecuador
\and
           Siena College, USA
          }

\abstract{%
  The CMS experiment at CERN has released research-quality data from particle collisions at the LHC since 2014. Almost all data from the first LHC run in 2010--2012 with the corresponding simulated samples are now in the public domain, and several scientific studies have been performed using these data. This paper summarizes the available data and tools, reviews the challenges in using them in research, and discusses measures to improve their usability.
}
\maketitle
\section{Introduction}
\label{intro}
The first data release of the CMS experiment was announced in 2014, bringing research-quality particle collision data into the public domain for the first time. Regular releases of CMS data have taken place ever since with modalities as defined in the data preservation, re-use and open access policy~\cite{data-policy}. As of 2021, more than 2 PB of data from the 2010--2012 run period are available to users external to the CMS collaboration, served through the CERN Open data portal (CODP)~\cite{CODP}.

The first research papers using CMS open data were published in 2017 on jet substructure studies~\cite{Tripathee:2017ybi, Larkoski:2017bvj}, and authors included valuable feedback and advice to the community. Further studies have been performed including searches for new particles~\cite{Cesarotti:2019nax, Lester:2019bso}, Standard Model analyses~\cite{Apyan:2019ybx}, and several studies on machine learning and methodology.

The usability of these data now and in the long-term future is the key factor when assessing the success, often requiring a delicate balance between constraints inherent to old data, the state-of-art tools at the moment, and the long-term vision.

In this note, Sec.~\ref{cms-od} briefly describes the CMS open data and associated products, available through the CERN open data portal. A typical analysis chain and the tools available for open data users are summarized in Sec.~\ref{use-od}. Feedback from CMS open data users and current limitations and challenges are discussed in Sec.~\ref{feedback}, and Sec.~\ref{improve} addresses the measures taken or foreseen to further improve the usability of CMS open data.

\section{CMS open data}
\label{cms-od}
Open data are released after an embargo period, which allows the collaboration to understand the detector performance and to exploit the scientific potential of these data.  In the latest update of the CMS data policy~\cite{data-policy}, the embargo period was set to six years after data taking. This is necessary for the time needed to reprocess the data with the best available knowledge before the release. The first release of each year's data consists of 50\% of the integrated luminosity recorded by the experiment, and the remaining data will be released within ten years, unless active analysis is still ongoing. However, the amount of open data will be
limited to 20\% of data with the similar centre-of-mass energy and collision type while such data
are still planned to be taken. This approach allows for a fairly prompt release of the data after a major reprocessing once the reconstruction has been optimised, but still guarantees that the collaboration will have the opportunity to complete the planned studies with the complete dataset first.
All data from 2010--2011 are now publicly available, as well as 50\% of proton-proton data from 2012.
 
\subsection{Data products}
\label{cms-od-data-products}
CMS releases a full reprocessing of data from each data-taking period in the Analysis Object Data (AOD) format, based on the ROOT framework~\cite{Brun:1997pa} and processed through CMS software CMSSW~\cite{cmssw}. The data are made available in the format and with the same data quality requirements that analyses of the CMS collaboration start from.

AOD is the main format used in CMS for Run-1 (2010--2012) data analysis. Starting from Run-2 (2015--2018), new reduced data formats called MiniAOD~\cite{Petrucciani:2015gjw} and NanoAOD~\cite{Rizzi:2019rsi} have been developed, and Run-2 data will be released in these slimmer formats. 

\vspace{4mm}

\noindent{\bf Collision data}

\noindent The collision data are stored in ``primary datasets'' based on the event content identified at the time of data taking. The dataset name is an indication of its physics content, and each dataset record lists the selection algorithms, i.e.\ the High-Level Trigger (HLT) streams, that were used to direct the data to that specific dataset. 

The typical data size of a single event in the AOD format is of the order of 0.1--0.3~MB depending on the year of the data taking and the beam conditions at that time. An event contains all data collected during one beam crossing, and it typically includes the hard-scattering event that has fired the selection algorithm, and several simultaneously happening soft-scattering events, so-called ``pile-up'' events. A primary dataset contains tens of millions of events, resulting in a total size of several terabytes per dataset. Each dataset consists of several files of the size of some GBs.

\vspace{4mm}

\noindent{\bf Simulated data}

\noindent The simulated datasets are generated by Monte Carlo generator programs, undergo detector simulation using CMSSW, and are subsequently processed into the same format as the collision data. During this processing chain, additional events are added on top of the simulated process to take into account the pile-up in the same beam crossing. The dataset names are identical to those used internally in CMS, and give an indication of the simulated process. Where possible, the exact parameters used in the event generation process, as well as in the further processing, are recorded. A keyword search has been implemented on the CERN open data portal to facilitate the search of simulated datasets of interest.

The average event size of simulated data is slightly larger than those of real collisions, because the information of the simulated process, the so-called ``Monte Carlo truth'', is included. The number of events in the simulated sample varies according to the process and depends on the study for which they were primarily intended.

\subsection{Software and associated information}
\label{cms-od-info}
The public data are accompanied by a compatible version of the CMSSW software and additional information necessary to perform a research-level physics analysis. The additional data products are needed in different steps of the analysis, for example for data selection, as correction factors to be applied to physics objects, or for evaluating the final measurable results in terms of cross sections. Example code and some specific guide pages are provided to explain and instruct the use of this associated information.

\vspace{4mm}

\noindent{\bf CMS software}

\noindent The CMS software, CMSSW, is open source and available on GitHub~\cite{cmssw}. It is also accessible to the CMS open data environment through the CernVM file system (CVMFS)~\cite{cvmfs}. This software is used for data taking, event reprocessing, and analysis, as well as for the generation of simulated events. 

A specific CMSSW version compatible with the data is needed. All production versions are available, but the older ones run only on by now outdated operating systems. They are made available as packaged computing environments, as discussed in Sec.~\ref{use-od}. For the available CMS open data, CMSSW versions on two different operating systems are required, Scientific Linux CERN 5 (SLC5) and 6 (SLC6) builds.

\vspace{4mm}

\noindent{\bf List of validated data}

\noindent For selecting properly validated events, each collision dataset has a corresponding list of validated runs, to be applied to each analysis to filter out data not having passed the data qualification criteria. This list is in JavaScript Object Notation (JSON) format, and tools are available in CMSSW to apply it as a filter in the analysis phase.

\vspace{4mm}

\noindent{\bf Information from the condition database}

\noindent In the analysis on the AOD format, values such as corrections for jet objects or information about the trigger selection are fetched from the condition database. This database contains non-event-related information (e.g. for alignment and calibration) and parameters for CMSSW software, and is accessed at run-time in the analysis jobs. The information needed at the AOD analysis level is a small fraction of the large database, but the complete collection of condition data is kept available to maintain the possibility for the full simulation, reconstruction and analysis chain. Access to the database is provided through CVMFS, which is made available to the CMS open data computing environment. 

\vspace{4mm}

\noindent{\bf Luminosity information}

\noindent The luminosity information is needed to express the measurements in terms of cross sections. Detailed luminosity listings by ``luminosity sections'', the smallest unit of data taking, are made available in CSV file format. Access is also provided to the original tool to ease the evaluation of luminosity for specific event selections.

\vspace{4mm}

\noindent{\bf Example code and topical guide pages}

\noindent Data releases are accompanied by example code repositories including some simplified analysis workflows and topical examples of data manipulation. In recent releases, some small-scale examples have been implemented as automated GitHub workflows.
These examples are searchable as software records on the portal, and, for ease of use, the code is also available from a linked GitHub repository.

Guide pages have been provided on the CERN open data portal on some specific technical topics, such as the trigger system, the condition database access, or the tools to access luminosity information. Some background information is covered on further guide pages, for example on dataset naming conventions or data processing chains.

\section{Using CMS open data}
\label{use-od}

Analysis of the CMS data is most commonly done in two steps: first, selecting events of interest and writing them to a new, smaller format, and second, analysing the selected events. Due to the experiment-specific data format, the first step will almost inevitably be done using the CMS software CMSSW in a computing environment compatible with the open data. For a realistic physics analysis, this step usually consists of hundreds of jobs, each taking several CPU hours. The analysts then have the option of either remaining in the open data environment, or moving their data out of the open data portal to their own computers for subsequent processing and optimization.

\begin{figure}
    \centering
    \includegraphics[width=\textwidth,bb=0 0 1344 576]{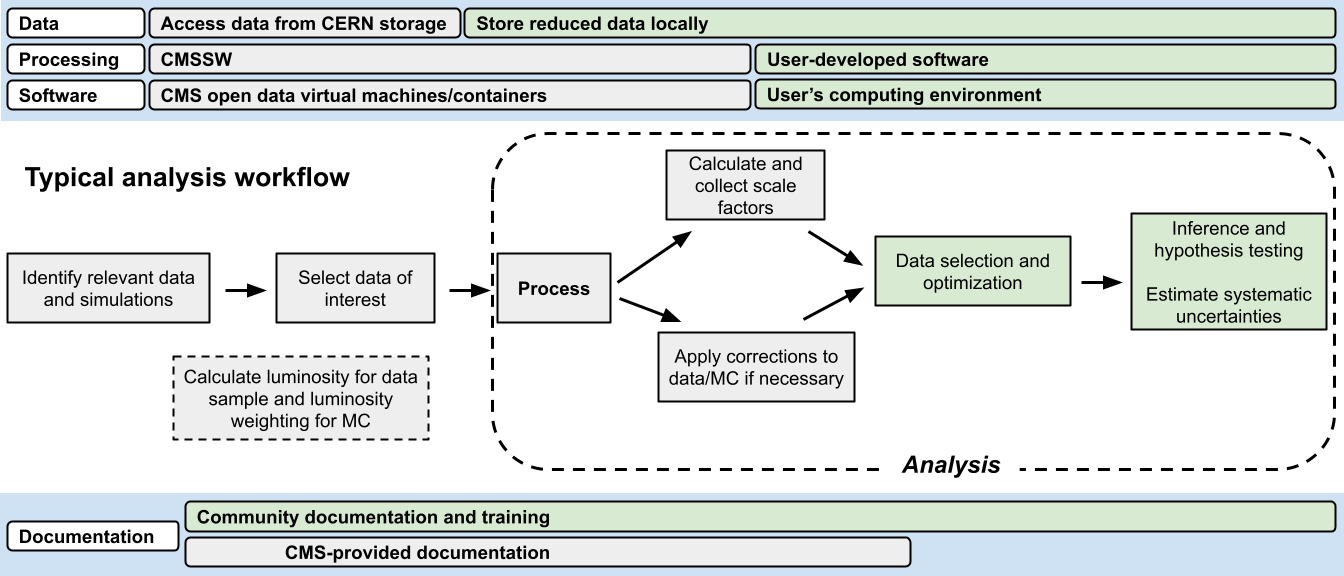}
    \caption{A coarse flowchart for a typical analysis. The upper part of the figure shows the hardware and software that an analyst might use for different stages of the analysis and the lower part of the figure shows where CMS’s documentation efforts span and end. Grey boxes indicate procedures, software, or hardware and storage are provided by the CERN or the CMS open data group. Light green boxes indicate steps where the procedures, software, or hardware and storage is left to the individual analyst.}
    \label{fig:flowchart}
\end{figure}

Fig.~\ref{fig:flowchart} provides a simplified flowchart of an analysis, as well as what
hardware, software, and documentation resources are needed/provided for the CMS open data.
In this section, we will elaborate on the details of what is provided and how they can
be accessed by users.

\subsection{Virtual machine images}
\label{use-vm}
Since the first release, the open data have been accompanied by virtual machine (VM) images based on the CernVM software appliance~\cite{cernvm}. These images can be run on any operating system through a hypervisor programs such as Oracle VM VirtualBox, and have the advantage of providing a quick first access to open data, and a full graphical user interface for the analysis program ROOT.

The virtual machine image is smaller than 20 MB and quick to download. It provides the base for the operating system compatible with the CMSSW version required for the open data as a singularity container within the VM, appearing to the user as a separate ``CMS shell''. The software environment is downloaded run-time and dependencies are accessed through CVMFS, which is mounted to the VM. After the download of the necessary software, the VM file size on the host machine is of order of GBs.

While VMs can be used on batch compute farms, for an open data user without a strong support of a local IT department, scaling up from the first quick data access to a computing setup adequate for a full analysis is a challenge. This has resulted in open data analyses being performed on a few desktop computers, running over weeks.

\subsection{Container images}
\label{use-cont}

With significant advances in the ease of use of operating system level virtualisation, mainly thanks to Docker~\cite{docker} and the Open Container Initiative (OCI)~\cite{oci}, software containers provide an alternative to the VM images. Container images are provided through the CERN Open Data portal and are available from Docker Hub and the CERN GitLab container registry. Different kinds of images are provided: light-weight images of the size of several hundred megabytes that only contain the base operating system needed to execute CMSSW and images that contain full CMSSW installations (several gigabytes). The latter images have the advantage of being independent of the underlying infrastructure, which means that they can be executed without network access and a CVMFS installation. Only the data and the corresponding condition database file need to be made available. The light-weight images require a CVMFS installation or mount to be available in order to be able to access CMSSW. Furthermore, container images that mount CVMFS from inside the container are also made available. However, this requires enhanced system privileges and reduces the isolation of the container from the host system and should therefore be used with care.

Since the container images use a layered file system, physics analysis code can be added on top of the base images without significantly increasing the storage space required, since the underlying layers will remain the same and can be reused independent of the analysis. This facilitates the complete preservation of physics analyses. The code can for example be added routinely using continuous integration system such as GitLab CI.

Container images can also be executed on batch compute systems. The container run-time used needs to support reading the OCI standard, which effectively all modern run-times do. High-throughput compute systems often only provide unprivileged (rootless) system access for the run-time for security reasons, which can e.g.\ be realised using Singularity or Podman. Container images provide a similar user experience as the VM images while also allowing the analysts to scale up to the level of a realistic physics analysis that requires a large amount of computing power.

\subsection{Data access}
\label{use-data}
The CMS open data files are stored at CERN. They can be accessed directly without prior download (streamed) using the XRootD protocol~\cite{xrootd} or downloaded using HTTP. For access via XRootD, file listings are attached to each data record on the portal and can directly be provided to CMSSW job configuration files. Streaming access has the advantage of not requiring local storage of several terabytes for data files. However, the access speed is affected by the network speed and decreases with increasing distance to CERN.

Single files of a dataset can be downloaded through HTTP from the portal web interface.  Command line tools are emerging for a direct download for any type of file from the open data portal, either through HTTP or XRootD protocol, but large-scale testing is still to be done. 

\subsection{Condition data}
\label{use-cond}
The condition database contains important information for the data processing as described in Sec.~\ref{cms-od-info}. The data are accessible to the CMS open data containers on CVMFS in SQLite format or from a Frontier database server. These databases vary in size, and range from 2 GB to 75 GB for collision data, depending on the collision type and the data-taking period. 

A download of large parts of the database at the start of the analysis job is launched even if only small part of it is needed in the actual analysis. This causes a significant overhead in the analysis.

\section{User feedback and challenges}
\label{feedback}
User questions and feature requests are received through the CERN open data portal support email, which has been active since the first release. More recently, a user forum~\cite{od-forum} has been set up, so that most common queries and their solutions can be easily available to new users. Furthermore, some authors have included their experience and recommendations in their papers reporting on their work on CMS open data~\cite{Tripathee:2017ybi, Apyan:2019ybx} and valuable feedback has been received in informal discussions. This section summarizes the user feedback, and challenges as perceived by the CMS open data team. The measures to address them are discussed in the following section.

\subsection{User feedback}
\label{feedback-user}
In their first study on CMS open data~\cite{Tripathee:2017ybi}, using the collision data from 2010 of the very first release, the authors reported challenges connected to the complexity of the CMSSW software and the resulting slow development cycle due to collision data streaming through XRootD and condition data download at the start of a job. Scattered documentation was considered as a problem, as well as the lack of validation examples for users to reproduce before setting up their own analysis. The availability of the full event information in the AOD format was appreciated for preservation purposes, but parsing the format for a specific study was necessary. The presence of superfluous data, e.g. those not having passed the validation criteria, or the events passing the trigger selection below the 100\% trigger efficiency threshold was considered to increase the data volume beyond necessary.

The benchmark study on Standard Model production cross sections~\cite{Apyan:2019ybx} using 2012 data reported that highest precision measurements were not achievable, given the limited available information on the detector calibration as well as the systematic uncertainties of relevant observables. Experimental uncertainties in the energy scale of particle jets and uncertainties in the lepton identification efficiencies were given as examples.

In both of these studies, the importance of providing a full analysis benchmark was emphasized. 

All the analysts who have used open data, however, have highlighted the importance of the research-quality data released by CMS and have rooted for the continuation of these efforts.  It is a common sentiment that further interest in these data by the external community will spur new particle-level studies, generate new search strategies and ideas, and expand certain measurements to unexplored phase-space regions, thereby propelling the progress of science.

\subsection{Data complexity}
\label{feedback-data}
The particle physics data are intrinsically complex. During the data processing, in the so-called reconstruction phase, ``physics object candidates'' and associated information needed to identify them are built from the detector signals. In terms of data structure, this results in a set of collections of objects of different sizes and types. The objects and their properties are defined as classes in the experiment-specific CMS software, which is built on top of the HEP-specific ROOT data structure.

Due to the large variety of different topics to be studied on these data, no single definition of a physics object exists. The AOD data contains multiple instances of an object, such as jet of particles reconstructed with different algorithms, or muons reconstructed taken into account either full detector information or only parts of it. Furthermore, a signal can be interpreted as several different object candidates, for example, a signal interpreted as a muon can also appear in the list of electron candidates. 

It is also to be taken into account that sometimes the identification of the actual particle having left the signal is impossible with the information available. This is the typical case for photons, and most of the signals interpreted as photons are due to neutral pions decaying into two photons, leaving almost equal signature in the detector. The object identification may also be affected by the presence of signals from the simultaneously occurring pile-up events.

Physics object identification is always needed in the data analysis, and the criteria are adapted balancing between the efficiency of selecting an object and the purity of the selected object. The choice depends on the type of analysis. Object identification and selection efficiencies, as well as eventual fake identification rates, are an important input in the final analysis result.

Final corrections and fine-tuning to the objects are often applied in the analysis phase. This happens because corrections and algorithms are developed at the same time as the first analysis of the data, and do not necessarily make their way to the final reprocessed data. Some of these corrections are available from the condition database, and some others as separate ``recipes'' to be applied to the objects.

Data-taking procedures add their own complication. An event is recorded if a trigger is fired at the time of data taking. There are hundreds of different trigger paths, and a single dataset consists of events passing one or more of them. In some cases, to enable collecting data for processes that occur often and otherwise would fill the entire data-taking bandwidth, triggers are prescaled so that only a predefined fraction of events passing the trigger selection is recorded. One event may also end up in two different datasets, for example, an event with two muon candidates may be included both in ``SingleMu'' and ``DoubleMu'' datasets, and this eventual overlap needs to be accounted for in the analysis. All this needs to be handled in the data analysis to properly scale the number of selected events with the cross sections being measured.

\subsection{Software complexity}
\label{feedback-sw}
The most common analysis workflow, as explained in Sec.~\ref{use-od}, consists of two or more steps, the first of which is done within the CMS open data environment using CMSSW. Mastering CMSSW, with more than 5 million lines of code, is a major challenge for external users. C++ skills are required, as well as a basic understanding of ROOT, which is the base of the data format. These are all very HEP-specific skills, and an investment in time on learning them is an additional hurdle for open data use.

While good documentation sources exist and the main CMS user guide WorkBook~\cite{workbook} is publicly available, coherent, versioned documentation is lacking. This is, in particular, a problem for open data users who need information on CMSSW versions that are no longer in use and who may not be able to access all CMS-internal information.
The available documentation rarely covers, in a concise form and ready to be applied on open data, the procedures at the analysis level. In addition, many of these procedures are fairly nonuniform, such as applying corrections and scale factors to different physics objects, as the methods depend on the working group in CMS providing them.

Of course, much of the analysis will be left up to the analysts. There is no one approach to optimizing the selection criteria and there is no one approach to how to extract a measurement from the data or set an upper-limit. While the CMS-open data group can point to community references or published analyses, it should be noted that the details of these steps are not codified in any ``official'' CMS documentation. 

\subsection{Scalability}

The amount of data to be accessed in a single analysis is easily of order of tens of terabytes or more. The bottleneck in the process is the first analysis step selecting the data of interest from the AOD (or AODSIM) files and writing them to a smaller analysis-specific format. For an analysis other than quick exploration this step requires a proper computing setup where jobs can be run automated in parallel. 

The remote data access can bring a considerable overhead, as each job, in addition to data assigned to a single job, will need to download the software base, and the condition data.
The current docker containers images (see Sec.~\ref{use-cont}) include the software but the time-consuming condition data access is still needed.

\subsection{Long-term usability}
The current approaches for the CMS Run-1 open data use so far rely on the availability of the CMS open data environment on which a compatible version of CMSSW can be run. The packaged computing environments such as VMs and docker containers images can preserve the needed environment. They are, however, not totally isolated from the rest of the computing world as data and software access is needed for the analysis job. Therefore, continuous maintenance of these environments is needed.

As discussed in Sec.~\ref{cms-od-data-products}, the future release of CMS Run-2 data will include data in NanoAOD format, which is a simplified data format and does not require CMSSW or a specific operating system to be accessed. While restricting the variety of analysis that can be performed, this format offers a good perspective for long-term usability. Run-1 data have not been reprocessed to these smaller formats, but work is ongoing to eventually offer an equivalent NanoAOD-type format for Run-1.

\section{Measures to improve usability}
\label{improve}
Taking into account the feedback received (Sec.~\ref{feedback-user}) by the researchers using open data, several measures have been (or are being) taken in order to improve usability.  The general strategy consists of agglutinating the scattered information into a well-written, centralized guide, procure frequent and regular training opportunities, prepare workflows as examples of well-defined, real analyses, and provide alternatives for the expansion of computing capabilities.

Obviously, all the actions required to make these improvements need non-negligible person-power.  CERN and CMS personnel involved, who are often devoting only a small part of their time to the open data adventure, are always looking for more efficient ways of implementing the ideas that will make the access to open data more robust. 

\subsection{Documentation and training}

There is already extensive documentation available on the usage of data taken with the CMS experiment.  Examples are the public pages of the CMS Twiki page, some topical guides on the CODP (see Sec.~\ref{cms-od-info}) and, above all, thousands of lines of code in the CMSSW Github repository.  In addition, there are several examples of analyses, tools and procedures that are implemented in CODP records.  However, as reported in Sec.~\ref{feedback-user}, it is not a trivial task to navigate through all these sources of information to conduct original research.

The problems with the CMS Twiki pages, despite their richness, have been already mentioned in Sec.~\ref{feedback-sw}. CODP guides and example records are also an important source of information.  However, they are not easy to find.  This is due to the fact that the platform is not designed to serve as a documentation repository, but rather as a dataset repository. Lastly, the CMSSW Github repository contains valuable information but it could be quite difficult, for a first user, to understand its structure and find the required information.

Many of the problems with these scattered sources of documentation can be and are being addressed by the implementation of a structured set of directions in a single CMS Open Data Guide~\cite{odguide} document, which is open to collaborative development.  Many of the issues arising due to the data and software complexity (see Sec. 4.2 and 4.3) are being tackled with this initiative.  The guide is not meant to repeat the information found in other sites but to hint analysts in a much orderly fashion.   This effort is complemented by the recent establishment of a user forum~\cite{od-forum}, where users have the opportunity to reach, many times directly, an internal CMS expert on a particular topic.

The organization of training events, like hands-on workshops, has been recently incorporated into the strategy for a better popularization of the usage of CMS open datasets.  Towards the end of 2020, the first workshop~\cite{odworkshop} on CMS open data took place remotely.  The usage of software lesson templates, as well as video tutorials and hands-on activities, resulted in a very positive feedback from the participants.  Brand new documentation, like improved luminosity estimations, was included in this event.

In general, the planning of frequent and regular workshops, around different sets of topics, will also improve the ability to address more questions that flow through the feedback lines.  Issues related to, e.g., object corrections or simplified baseline calibrations and uncertainties, will get the opportunity to be dealt with in a more pedagogical way, with better and more robust documentation. 

\subsection{Preserved workflows as examples}

Actual physics analysis workflows are crucial for the usability of open data. They provide examples that can be adapted to the respective use cases by the open data analysts. This, in particular, helps overcome some of the challenges inherent to the data and software complexity discussed in Sec.~\ref{feedback}.

The form in which open data workflows are currently preserved varies. All workflows have a dedicated record on the CERN open data portal, providing a general overview . While the code is preserved as part of the record, it is furthermore made available through GitHub repositories linked from the record. These repositories contain more detailed workflow descriptions and instructions on how to execute them. In several cases, examples of increasing complexity, in particular with respect to the required computing power, are provided, e.g.\ the ``Higgs-to-four-lepton analysis example using 2011--2012 data''~\cite{cms-od-h4l}. This example covers four different levels of complexity with overall execution times ranging from minutes to weeks (in case of no parallel execution).

Another example~\cite{cms-od-wunschtuple}, which has mainly been developed for outreach, education, and software development, demonstrates how one can obtain a simplified, ``ntuple'' data analysis format, analysis of which does not require CMSSW, but only ROOT, and is also independent of the original Linux distribution needed to run CMSSW. A subsequent analysis example is also provided~\cite{cms-od-htau}. It has to be noted, however, that the ntuple is only suited for the given analysis example and extending the format to cover more use cases requires CMS expertise.

When providing analysis examples, their initial validation is very important to ensure they work as desired. Furthermore, they have to be executed on a regular basis to confirm they continue to work. For example, external dependencies such as the container registry or the condition database could change and require changes to the analysis code and/or setup. This validation is mostly carried out using continuous integration features provided by both the GitHub~\cite{github-actions} and GitLab~\cite{gitlab-cicd} platforms running on individual test files for individual analysis steps. Several issues have been identified and resolved in this way over the last year.

In addition to testing individual analysis steps on a regular basis, also the overall workflows need to be exercised. For this purpose, a dedicated Kubernetes~\cite{Kubernetes} cluster has been set up at CERN, with which the software containers can be orchestrated. The workflow engine used to connect the workflow steps is ``Argo Workflows''~\cite{argo-workflows}. This engine can also be used directly on the GitHub and GitLab platforms by using a single-node Kubernetes cluster created with ``minikube''~\cite{minikube}. Furthermore, some workflows are also implemented and executable on REANA~\cite{reana}, a platform for reproducible research data analysis developed at CERN.

\subsection{Data analysis in cloud environments}

One of the biggest challenges when trying to run CMS open data analyses at scale is the required computing power, see also Sec.~\ref{use-cont}. The software execution environment can be largely hidden away by the use of software containers (cf.\ Sec.~\ref{use-cont}). If a computing cluster is available to the analysts, they should generally be able to perform open data analyses on it. The workflow, however, would most likely have to be reimplemented based on the workflow engines available on the respective cluster.

In case there is no local computing cluster available, a large number of cloud compute products are publicly available. These allow for the dynamic scaling of the required resources, which is particularly important because one is generally only charged for the resources provisioned. The analysts can e.g.\ develop their analyses on a single machine and then scale up to several machines to execute the full analysis workflow over all datasets. Nevertheless, particular care is needed to ensure that the resources are scaled down again once they are not needed anymore as costs can quickly go into the hundreds of Swiss francs.

When making use of public cloud offerings, Kubernetes clusters provide a very flexible environment. Scaling of resources based on load happens automatically, and Kubernetes also provides an application programming interface (API) that allow for the creation of batch jobs. Kubernetes, however, has a steep learning curve, and more work is needed to simplify the overall setup of a Kubernetes cluster for CMS open data analysis. Nevertheless, software solutions such as Argo Workflows and REANA significantly simplify the execution of analysis workflows on Kubernetes clusters.

While data traffic into the public compute clusters is mostly free of charge, streaming the data via XRootD (see Sec.~\ref{use-data}) can significantly slow down the analysis. It is often better to download the required datasets to the computing platform first. Further performance improvements can be achieved by using a local container registry for the container images and making use of CVMFS with a local cache server.

\section{Outlook}
Even with the particular challenges presented by collider experiments, the CMS collaboration has demonstrated that making datasets available to external users is not only possible, but also yields new physics results. There is excitement in the theory community surrounding this opportunity for collaboration with experimentalists in a totally new fashion. The recent theory workshop has provided concrete feedback as to how CMS can facilitate the access and analysis of these datasets which will allow CMS to provide targeted improvements to the open data efforts. This work will no doubt benefit the broader HEP community as we work together to produce the greatest scientific output from these rich datasets.

\bibliography{refs}

\end{document}